\documentclass[aps,nofootinbib,showpacs,showkeys,preprint]{revtex4}

\usepackage{epsf,epsfig,subfigure,graphicx,amsmath,amssymb}
\usepackage{color}

\def\lsim{\mathrel{\rlap{\lower4pt\hbox{\hskip1pt$\sim$}}
    \raise1pt\hbox{$<$}}}         %less than or approx. symbol
\def\gsim{\mathrel{\rlap{\lower4pt\hbox{\hskip1pt$\sim$}}
    \raise1pt\hbox{$>$}}}         %greater than or approx. symbol

\begin{document}

\title{\bf X-ray line signal from 7 keV axino dark matter decay}

\author{
Kyoungchul Kong$^{(a)}$\footnote{email: kckong@ku.edu},
Jong-Chul Park$^{(a, b)}$\footnote{email: log1079@gmail.com}, and
Seong Chan Park$^{(b)}$\footnote{email: s.park@skku.edu}
}
\affiliation{$^{(a)}$
Department of Physics and Astronomy, University of Kansas, Lawrence, KS 66045, USA
\\
$^{(b)}$ Department of Physics, Sungkyunkwan University, Suwon 440-746, Republic of Korea
}

\begin{abstract}
Recently a weak X-ray emission around $E_\gamma \simeq 3.5$ keV was detected in the Andromeda galaxy and various galaxy clusters including the Perseus galaxy cluster but its source has been unidentified. Axino, the superpartner of axion, with a mass $ 2E_\gamma$ is suggested as a possible origin of the line with R-parity violating decay into photon and neutrino.
Moreover, most of parameter space is consistent with recent observation by the BICEP2 experiment.
\end{abstract}

% \pacs{12.90.+b, 14.80.Ly, 95.35.+d}

 %95.35.+d DM
 %98.62.Gq galactic halos
 %98.70.Rz gamma ray sources
 %95.85.Ry neutrino muon and other elem. particles,
 %11.30.Ly: other internal and higher symmetries
 %12.60.Cn: extension of EW sector
 %12.90.+b: Miscellaneous models
 %14.70.Pw: Other gauge bosons

\keywords{X-ray, Neutrino, Dark matter, keV events, Axino, BICEP2}

\maketitle

\section{Introduction}

The recent two independent analyses~\cite{7keV1,7keV2} based on X-ray observation data show an emission line at $E \simeq 3.5$ keV in the spectra coming from various galaxy clusters and the Andromeda galaxy. The observation is statistically significant  ($ \sim 3\sigma-4 \sigma$) and more importantly is quite consistent with the location of the line in energy spectra and the signal flux. The observed flux and the best fit energy are
\begin{eqnarray}
{\Phi}^{\rm MOS}_\gamma &=& 4.0^{+0.8}_{-0.8} \times 10^{-6}\, {\rm photons ~cm^{-2} ~s^{-1}}\,,\\
E^{\rm MOS}_\gamma &=& 3.57 \pm 0.02\, {\rm keV}\,,
\end{eqnarray}
where we take the values from the XMM-Newton MOS spectra, and the results from the PN observations are similar~\cite{7keV1} and consistent with the measured values in the other analysis~\cite{7keV2}.

No source of X-ray line including atomic transition in thermal plasma is known at this energy, which indicates that
the observed line may suggest the existence of a new source.
It would be tantalizing if a dark matter (DM) provided a possible source for the line signal.
Indeed,  a decaying DM of a mass $m_{\rm DM} \simeq 2 E_\gamma \simeq 7$ keV and a lifetime $\tau_{{\rm DM} \to \gamma X} \simeq 10^{28}\, {\rm s}$ is immediately suggested to explain the observed line~\cite{7keV1,7keV2}. An annihilating DM of a mass $m_{\rm DM} \simeq E_\gamma \simeq 3.5$ keV and an annihilation cross section $\langle \sigma v\rangle_{{\rm 2 DM} \to \gamma X} \sim 2 \Gamma_{\chi}/n_\chi \sim (10^{-31}-10^{-33})~{\rm cm^3 ~s^{-1}}$ can also account for the signal, where $n_\chi =\rho_\chi/m_\chi \sim (10^3-10^5)~{\rm cm^{-3}}$ is the DM number density of galaxy clusters.
However, the realization of such an annihilating DM is very challenging since the
corresponding annihilation cross section is too small compared to a typical value for a thermal WIMP (weakly interacting massive particle) DM.
Other annihilation channels are also limited due to the small DM mass. Hereafter we will focus on a decaying DM model.
Possible DM candidates such as a sterile neutrino and a long lived axion have been suggested
as an explanation of this signal \cite{Ishida:2014dlp,Finkbeiner:2014sja,Higaki:2014zua,Jaeckel:2014qea, Lee:2014xua,Abazajian,Krall}.\footnote{For the cases of decaying sterile neutrino and gravitino warm dark matter, the authors of Ref.~\cite{Abazajian:2001vt} estimated expected x-ray fluxes from galaxy clusters and field galaxies.}
To explain the 3.5 keV line with 7 keV axion DM \cite{Higaki:2014zua,Jaeckel:2014qea}, the required axion decay constant $f_a \simeq 10^{14-15}$ GeV, which is much larger than the conventional values preferred by most axion models~\cite{Axion,AxionReview}.

In this letter, as an alternative, we examine axino ($\tilde{a}$) as a dark matter candidate and show how axino can fit the observed data.
With an axion~\cite{Axion,AxionReview} as a solution of the strong CP problem,  a light axino with a mass $m_{\tilde{a}}\sim \frac{M_{SUSY}^2}{f_a} \sim 7~{\rm keV}$  is an excellent DM candidate in supersymmetric models \cite{Covi:1999ty, Covi:2001nw, Covi:2009pq}.
Moreover, it has been shown that axino in the preferred mass range can be a warm dark matter (WDM) satisfying the relic density constraint \cite{Covi:2009pq} through thermal production via thermal scatterings and/or non-thermal production via out-of-equilibrium decays.

WDM is known to provide a solution to the small scale conflict between the observations and the N-body simulations with cold dark matter (CDM),
where the overproduction of galactic substructures~\cite{Moore:1999nt}, local groups~\cite{Zavala:2009ms}, and local voids~\cite{Tikhonov:2008ss} compared to the observations has been found. A lower limit on WDM mass is $m_{\rm WDM} > 3.3$ keV from the recent high red-shift Lyman-$\alpha$ forest data~\cite{Viel:2013fqw}.
The small scale behaviors of WDM with $ m_{\tilde{a}} \gtrsim 4-5$ keV are not so different from those of CDM \cite{Maccio':2012uh,Schneider:2013wwa}. Consequently, the 7 keV axino can alleviate a little of the small scale problems of CDM.

\section{Axino dark matter with R-parity violation}

The axino can be a good DM candidate even in the presence of R-parity violation.
The decay channel to neutralino and photon, $\tilde{a} \to \tilde{\chi}_0 \gamma$, is kinematically closed with the heavier mass of neutralino, $m_{\tilde{\chi}^0} > m_{\tilde{a}}$, thus the axino mainly decays to the standard model particles through R-parity violating interactions. The decay width however is strongly suppressed by a high Peccei-Quinn symmetry breaking scale, $f_a$, and also the small R-parity violation so that the resultant lifetime can be long enough.

A bilinear type of the R-parity violation is considered as a simple model \cite{Hall:1983id},\footnote{A similar model has been used to explain 130 GeV gamma-ray line signal from the galactic center region \cite{Endo:2013si}.} which is described by the following superpotential,
\begin{eqnarray}
W_{\not{R}} = \mu_i L_i H_u\,,
\end{eqnarray}
where $L_i$ and $H_u$ are respectively the lepton doublet and the up-type Higgs superfields and the index $i = \{1,2,3\}$ runs over generations. With these R-parity violating terms, the axino decays into a photon and a neutrino, and the decay rate is given by \cite{Covi:2009pq,Endo:2013si},
\begin{eqnarray}\label{decay}
\Gamma_{\widetilde{a} \to \gamma \nu_i} = \frac{m_{\widetilde{a}}^3}{128 \pi^3 f_a^2}\, \alpha_{em}^2 C_{a\gamma\gamma}^2 |U_{\nu_i\widetilde{\gamma}}|^2
\end{eqnarray}
with the neutrino-photino mixing parameter $U_{\nu_i\widetilde{\gamma}}$,
\begin{eqnarray}
U_{\nu_i\widetilde{\gamma}} \simeq \xi_i \frac{\sqrt{2}s_W M_Z}{M_1}\,,
\end{eqnarray}
where $\xi_i = \langle \widetilde{\nu}_i \rangle / v$ with the vacuum expectation values (VEV) of sneutrinos $\langle \widetilde{\nu}_i \rangle$ and Higgs $v = 246$ GeV. $M_Z$ is the mass of the standard model $Z$ boson, $M_1$ is the U(1)$_Y$ gaugino mass, and $s_W=\sin\theta_W$ where $\theta_W$ is the Weinberg angle.
$C_{a\gamma\gamma}$ is a model dependent constant of order unity, which is normalized as
\begin{eqnarray}
\mathcal{L} = \frac{\alpha_{em}}{8\pi f_a}\, C_{a\gamma\gamma}\, a F_{\mu\nu} \widetilde{F}^{\mu\nu}\,,
\end{eqnarray}
where $\alpha_{em} = e^2/4\pi$, $a$ is the axion field, and $F_{\mu\nu}$ and $\widetilde{F}^{\mu\nu}$ are the electromagnetic field strength and its dual respectively.

The lifetime of axino is conveniently given as
\begin{eqnarray}\label{axinolifetime}
\tau_{\widetilde{a} \to \gamma \nu} \simeq 10^{28} {\rm s}\, \left( \frac{C_{\rm eff}}{10^{-6}} \right)^{-2} \left( \frac{m_{\widetilde{a}}}{7 ~{\rm keV}} \right)^{-3} \left( \frac{f_a}{3\times 10^8 {\rm \, GeV}} \right)^{2}\,,
\end{eqnarray}
where
$C_{\rm eff} \equiv C_{a\gamma\gamma} \sum_i |U_{\nu_i\widetilde{\gamma}}| \simeq C_{a\gamma\gamma}\, \xi\, \frac{\sqrt{2}s_W M_Z}{M_1}$ with $\xi \equiv \sqrt{\sum_i |\xi_i|^2}$.
Thus, the axino DM with 7 keV mass can be a good source for the 3.5 keV X-ray line signal with a reasonable choice of parameters: $C_{\rm eff}\simeq 10^{-6}$ and $f_a\simeq 3\times 10^8$ GeV.

The sneutrino VEVs $\langle \widetilde{\nu}_i \rangle$ induce the mixing between the leptons and the neutralinos generating neutrino masses at the tree-level.
From the upper bound on the neutrino mass $m_\nu \equiv \sum_i m_{\nu_i} < 0.23$ eV ($95\%$ C.L., \textit{Planck}+WP+highL+BAO) \cite{Planck}, we estimate the size of $\xi$ \cite{Chun:2004mu}:
\begin{eqnarray}\label{NuMass}
\xi \lesssim \frac{1.1}{\cos\beta} \times 10^{-6} \left( \frac{M_N}{M_Z} \right)^{1/2} \left( \frac{m_{\nu}}{0.23~{\rm eV}} \right)^{1/2}\,,
\end{eqnarray}
where $M_N = M_1 M_2 / (c_W^2 M_1 + s_W^2 M_2) + M_Z^2 s_{2\beta}/\mu$ with the gaugino masses $M_1,M_2$ and $\mu$ parameters as in Ref.~\cite{Choi:1999tq}. With a choice of parameters, $M_1=M_2= M_{1/2}\simeq (1-10)~{\rm TeV}$ and $\mu > M_Z$, we get $\xi \lesssim (0.5-12)\times 10^{-5}$ with $\tan\beta= (1-10)$.  Thus, $C_{\rm eff}\simeq 10^{-6}$ is obtainable with $C_{a\gamma\gamma} \sim \mathcal{O}(1)$ and $M_1 \sim 1$ TeV, which is natural.

\begin{figure}[t]
\begin{center}
\includegraphics[width=0.80\linewidth]{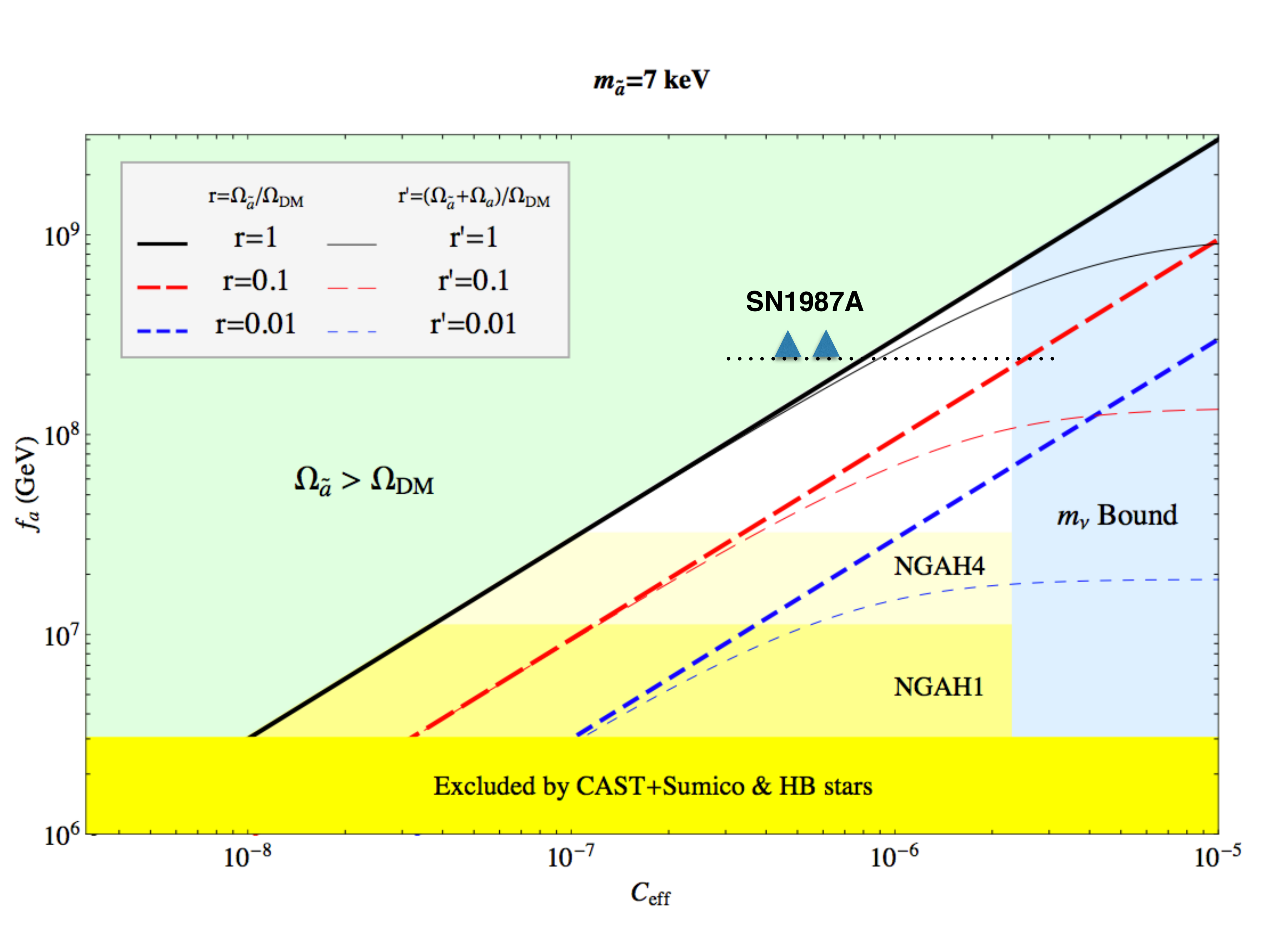}%Ceff-fa_BICEP.pdf}
\end{center}
\vspace*{-0.7cm}
\caption{Parameter space that is consistent with the 3.5 keV line signal in the $C_{\rm eff}-f_a$ plane.
The three thick-straight lines represent the values of $C_{\rm eff}$ and $f_a$ to fit the required lifetime for different values of $r$, $r=\frac{\Omega_{\widetilde{a}}}{\Omega_{\rm DM}} = 1, 0.1$, and $0.01$, respectively.
The three thin curves respectively correspond to $r' = \frac{\Omega_{\widetilde{a}}+\Omega_a}{\Omega_{\rm DM}} = 1, 0.1$, and $0.01$ for $\alpha^{\rm dec} =186$.
The upper (green), lower (yellow), and right (blue) shaded regions are excluded by the DM relic density, axion-like particle searches and astrophysical observations, and the neutrino mass limit, respectively.
Conservative projected limits from NGAH are shown as the (light-yellow) shaded regions.
A dotted horizontal line with arrows represents a model dependent bound from the SN1987A.}
\label{Fig1}
\end{figure}

In general, WDM is believed to comprise some portion of the observed DM relic abundance with CDM as a dominant source~\cite{Maccio':2012uh,Anderhalden:2012qt}. Taking this fact into account, we generalize our analysis by introducing a parameter,
\begin{eqnarray}
r = \frac{\Omega_{\widetilde{a}}}{\Omega_{\rm DM}}\,,
\end{eqnarray}
which describes the WDM portion in the total DM amount. With a suppressed value  $0 \leq r\leq 1$ in general,  the required lifetime for the observed flux is scaled linearly as
\begin{eqnarray}\label{requiredlife}
\tau_{{\rm DM} \to \gamma X} = \Gamma_{{\rm DM} \to \gamma X}^{-1} \simeq r \times 10^{28}\, {\rm s}\,,
\end{eqnarray}
because the expected X-ray flux is proportional to the density of WDM.
Comparing Eqs. (\ref{axinolifetime}) and (\ref{requiredlife}), one can easily find the needed values of parameters, $C_{\rm eff}$ and $f_a$, for a given axino DM fraction $r$.
In Figure~\ref{Fig1}, we show the parameter space that is consistent with the $3.5$ keV line signal in the $C_{\rm eff}-f_a$ plane for the representative values of $r=1, 0.1$, and 0.01. The upper left region of the thick solid (black) line is excluded since the required axino relic density $\Omega_{\widetilde{a}}$ is larger than the observed DM density $\Omega_{\rm DM}$.

%
%%%%%%%%%%%%%%%%%%%%%%%%%%%%%%%%%%%%%%%%%%%%%%%%%%%%%%%%%%%%
%%%%%%%%%%%%%%%%%%%%%%%%%%%%%%%%%%%%%%%%%%%%%%%%%%%%%%%%%%%%
%

Axions can be copiously produced through the $a \gamma \gamma$ interaction in the Sun,
which have been searched by axion telescopes such as CAST and Sumico.
In addition, the $a \gamma \gamma$ interaction can induce exotic cooling mechanisms in the cores of stars and thus affect stellar evolution,
which is constrained by observations of Horizontal Branch (HB) stars in globular clusters.
The lower (yellow) shaded region is constrained by axion-like particle search experiments and astrophysical observations~\cite{Jaeckel:2010ni}.
A future solar axion telescope, NGAH~\cite{Irastorza:2011gs}, provides  projected limits stronger than current bounds from CAST,
which are shown as (light-yellow) shaded regions in the figure.
We use the limits for $m_a > \mathcal{O}(10^{-2})\, {\rm eV}$ to be conservative.
They are $2-3$ times more stringent for a lower mass, $m_a <\mathcal{O}(10^{-2})\, {\rm eV}$.
With a representative parameter set of $\tan\beta = 10$,  $M_{1/2} = 1$ TeV, and $C_{a\gamma\gamma} = 1$,
we obtain a limit $C_{\rm eff} \lesssim 2.3 \times 10^{-6}$ with $m_{\nu} < 0.23$ eV, which appears as the right (blue) shaded region.
Another potentially important bound arises from the SN1987A. If the axion is hadronic, it can contribute to the emission of the energy from the SN1987A and provides a model dependent bound, $f_a \gtrsim 3.7\sqrt{F} \times (T/30{\rm MeV})^2 \times 10^8 $ GeV \cite{Raffelt:2006cw}.  With a temperature $T\simeq 30$ MeV and the axion absorption rate, $\Gamma_a/T \in (1,10)$, we find $F\in (0.46,1.35)$  or $f_a \gsim (2.5-4.3)\times 10^8$ GeV, which still allows the preferred value $f_a =3\times 10^8$ GeV.
However, these limits remain fairly rough estimates \cite{Raffelt:2006cw}, and
to completely rule out this scenario, one would require more robust experimental bounds such as
a future solar axion telescope, NGAH, as shown in Figure~\ref{Fig1}.

Finally the phenomenology of axion dark matter depends on the value of the Hubble expansion rate $H_I$ at the end of inflation, which can be determined by the tensor-to-scalar ratio and other CMB data \cite{Visinelli:2014twa}.
Recent observation of the primordial B-mode polarization by BICEP2 Collaboration \cite{BICEP2} favors an axion scenario with $f_a < \frac{H_I}{2\pi}$, which affects our discussion.
In such a scenario, the total axion energy density is given by
\begin{equation}
\Omega_a h^2 = 2.07 \times 10^{-14}\, ( \alpha^{\rm dec} + 1)\,  f_a^{7/6} \,,
\end{equation}
where $\alpha^{\rm dec}$ is the fractional contribution to the axion density from decays of axionic topological defects (see Ref. \cite{Visinelli:2014twa} and the references therein).
The axion abundance slightly modifies the allowed parameter space given by the three thick-straight lines.
It favors a lower $f_a$ for a higher $C_{\rm eff}$, while there is almost no effect for a lower $C_{\rm eff}$.
Modified bounds are shown as thin curves for the representative values of $r'= (\Omega_{\tilde a}+\Omega_a)/\Omega_{{\rm DM}}=1,0.1,$ and 0.01
with $\alpha^{\rm dec}=186$ in Figure~\ref{Fig1}.
The new bounds approach the straight lines with a smaller value of $\alpha^{\rm dec}$.

\section{Phenomenological implications}

The axino dark matter scenario resembles gravitino dark matter models.
Its collider signatures depend strongly on what the next-to-lightest supersymmetric particle is, providing a variety of possibilities.
Other signatures may arise as well due to the bilinear R-parity violations in this particular scenario.
However unfortunately it would be difficult to discriminate this model from other ``look-alike'' models.

What is more interesting is neutrinos from the axino decay.
As a byproduct of the axino decay, neutrinos with $E_\nu \sim 3.5$ keV are also expected as the same amount of 3.5 keV X-rays flux, $\sim 4 \times 10^{-6}\, {\rm cm}^{-2}\, {\rm s}^{-1}$. It would be challenging, if not impossible, to detect these neutrinos, because a large neutrino background is expected from various processes in the Sun as well as in the Earth: the solar fusion, thermal processes in the solar core and the terrestrial neutrinos by the natural radioactivity of the Earth. The expected neutrino background flux density at $E_\nu \sim 3.5$ keV is at the level of $10^8\, {\rm cm}^{-2}\, {\rm s}^{-1} {\rm MeV^{-1}}$ \cite{Haxton:2000xb}, which is much larger than that from the axino decay.
Moreover, there is currently no easy way to detect $\mathcal{O}({\rm keV})$ neutrinos since even electron is too heavy to be scattered off by keV neutrinos. In the future, bolometric detectors may be able to measure temperature changes by scatterings.
It will be extremely important to measure the neutrino events at $E_\nu = E_\gamma \sim 3.5 ~{\rm keV}$ to confirm the model. We would encourage the experimentalist to overcome the background with new neutrino detectors such as a directional detector in the future.

\section{Conclusion}

Recent observation of $E_\gamma \simeq 3.5$ keV X-ray line in galaxy clusters and Andromeda galaxy opens a new way to see dark matter particle in a light mass domain: $m_{DM} \simeq 3.5$ keV for an annihilating dark matter and $7$ keV for a decaying dark matter. In general, a long lived particle, which produces enough number of photons, could be a good candidate of the source of X-rays. In this letter, we studied the axino decay through the bilinear R-parity violating interaction. We found that the parameter space which fits the observed line is naturally compatible with most axion models as well as
recent observation by the BICEP2 experiment.
Observation of a neutrino line at the same energy, $E_\nu =E_\gamma=m_{\tilde{a}}/2$, as in the X-ray data, corroborates the
axino DM scenario.

\vspace{0.5 cm}
\begin{acknowledgements}
K.K. is supported by the U.S. DOE under Grant No. DE-FG02-12ER41809 and by the University of Kansas General Research Fund allocation 2301566.
J.C.P. is supported by Basic Science Research Program through the National Research Foundation of Korea funded by the Ministry of Education (NRF-2013R1A1A2061561).
S.C.P. is supported by Basic Science Research Program through the
National Research Foundation of Korea funded by the Ministry of
Science, ICT $\&$ Future planning (2011-0010294) and the Ministry of Education (2011-0010294, 2011-0029758, and NRF-2013R1A1A2064120).
\end{acknowledgements}


\begin{thebibliography}{99}



%\cite{7keV1}
\bibitem{7keV1}
  E.~Bulbul, M.~Markevitch, A.~Foster, R.~K.~Smith, M.~Loewenstein and S.~W.~Randall,
  %``Detection of An Unidentified Emission Line in the Stacked X-ray spectrum of Galaxy Clusters,''
  arXiv:1402.2301 [astro-ph.CO].

%\cite{7keV2}
\bibitem{7keV2}
  A.~Boyarsky, O.~Ruchayskiy, D.~Iakubovskyi and J.~Franse,
  %``An unidentified line in X-ray spectra of the Andromeda galaxy and Perseus galaxy cluster,''
  arXiv:1402.4119 [astro-ph.CO].




%%%%%%%%%%%%%%%%%%%%%%%% models for 3.5 kev line %%%%%%%%%%%%%%%

%\cite{Ishida:2014dlp}
\bibitem{Ishida:2014dlp}
  H.~Ishida, K.~S.~Jeong and F.~Takahashi,
  %``7 keV sterile neutrino dark matter from split flavor mechanism,''
  arXiv:1402.5837 [hep-ph].

%\cite{Finkbeiner:2014sja}
\bibitem{Finkbeiner:2014sja}
  D.~P.~Finkbeiner and N.~Weiner,
  %``An X-Ray Line from eXciting Dark Matter,''
  arXiv:1402.6671 [hep-ph].

%\cite{Higaki:2014zua}
\bibitem{Higaki:2014zua}
  T.~Higaki, K.~S.~Jeong and F.~Takahashi,
  %``The 7 keV axion dark matter and the X-ray line signal,''
  arXiv:1402.6965 [hep-ph].

%\cite{Jaeckel:2014qea}
\bibitem{Jaeckel:2014qea}
  J.~Jaeckel, J.~Redondo and A.~Ringwald,
  %``A 3.55 keV hint for decaying axion-like particle dark matter,''
  arXiv:1402.7335 [hep-ph].


%\cite{Lee:2014xua}
\bibitem{Lee:2014xua}
  H.~M.~Lee, S.~C.~Park and W.~-I.~Park,
  %``Cluster X-ray line at $3.5\,{\rm keV}$ from axion-like dark matter,''
  arXiv:1403.0865 [astro-ph.CO].
  %%CITATION = ARXIV:1403.0865;%%

%\cite{Abazajian}
\bibitem{Abazajian}
  K.~N.~Abazajian,
  %``Resonantly-Produced 7 keV Sterile Neutrino Dark Matter Models and the %Properties of Milky Way Satellites,''
  arXiv:1403.0954 [astro-ph.CO].

%\cite{Krall}
\bibitem{Krall}
  R.~Krall, M. Reece and T. Roxlo,
  %``Effective field theory and keV lines from dark matter,''
  arXiv:1403.1240 [hep-ph].

%%%%%%%%%%%%%%%%%%%%%%%%%%%%%%%%%%%%%%%%%%%%%%%%%%%%%%%%%%%%%%%%%%


%\cite{Abazajian:2001vt}
\bibitem{Abazajian:2001vt}
  K.~Abazajian, G.~M.~Fuller and W.~H.~Tucker,
  %``Direct detection of warm dark matter in the X-ray,''
  Astrophys.\ J.\  {\bf 562}, 593 (2001)
  [astro-ph/0106002].



%%%%%%%%%%%%%%%%%%%%%%%% Strong CP & axion %%%%%%%%%%%%%%%%%%%%%%%
%\cite{Axion}
\bibitem{Axion}
  J.~E.~Kim,
  %``Weak Interaction Singlet and Strong CP Invariance,''
  Phys.\ Rev.\ Lett.\  {\bf 43}, 103 (1979);
%
%%\cite{Shifman:1979if}
%\bibitem{Shifman:1979if}
  M.~A.~Shifman, A.~I.~Vainshtein and V.~I.~Zakharov,
  %``Can Confinement Ensure Natural CP Invariance of Strong Interactions?,''
  Nucl.\ Phys.\ B {\bf 166}, 493 (1980).

%\cite{AxionReview}
\bibitem{AxionReview}
For reviews, see J.~E.~Kim,
  %``Light Pseudoscalars, Particle Physics and Cosmology,''
  Phys.\ Rept.\  {\bf 150}, 1 (1987);
%
%%\cite{Kim:2008hd}
%\bibitem{Kim:2008hd}
  J.~E.~Kim and G.~Carosi,
  %``Axions and the Strong CP Problem,''
  Rev.\ Mod.\ Phys.\  {\bf 82}, 557 (2010)
  [arXiv:0807.3125 [hep-ph]].

%%%%%%%%%%%%%%%%%%%%%%%%%%%%%%%%%%%%%%%%%%%%%%%%%%%%%%%%%%%%%



%%%%%%%%%%%%%%%%%%%%%% axino DM %%%%%%%%%%%%%%%%%%%%%%%%%%%%%%%%%%%%

%\cite{Covi:1999ty}
\bibitem{Covi:1999ty}
  L.~Covi, J.~E.~Kim and L.~Roszkowski,
  %``Axinos as cold dark matter,''
  Phys.\ Rev.\ Lett.\  {\bf 82}, 4180 (1999)
  [hep-ph/9905212].

%\cite{Covi:2001nw}
\bibitem{Covi:2001nw}
  L.~Covi, H.~-B.~Kim, J.~E.~Kim and L.~Roszkowski,
  %``Axinos as dark matter,''
  JHEP {\bf 0105}, 033 (2001)
  [hep-ph/0101009].

%\cite{Covi:2009pq}
\bibitem{Covi:2009pq}
  L.~Covi and J.~E.~Kim,
  %``Axinos as Dark Matter Particles,''
  New J.\ Phys.\  {\bf 11}, 105003 (2009)
  [arXiv:0902.0769 [astro-ph.CO]].

%%%%%%%%%%%%%%%%%%%%%%%%%%%%%%%%%%%%%%%%%%%%%%%%%%%%%%%%%



%%%%%%%%%%%%%%%%%%%%%% small scale problems of CDM %%%%%%%

%\cite{Moore:1999nt}
\bibitem{Moore:1999nt}
  B.~Moore, S.~Ghigna, F.~Governato, G.~Lake, T.~R.~Quinn, J.~Stadel and P.~Tozzi,
  %``Dark matter substructure within galactic halos,''
  Astrophys.\ J.\  {\bf 524}, L19 (1999)
  [astro-ph/9907411];
%%\cite{Klypin:1999uc}
%\bibitem{Klypin:1999uc}
  A.~A.~Klypin, A.~V.~Kravtsov, O.~Valenzuela and F.~Prada,
  %``Where are the missing Galactic satellites?,''
  Astrophys.\ J.\  {\bf 522}, 82 (1999)
  [astro-ph/9901240].


%\cite{Zavala:2009ms}
\bibitem{Zavala:2009ms}
  J.~Zavala, Y.~P.~Jing, A.~Faltenbacher, G.~Yepes, Y.~Hoffman, S.~Gottlober and B.~Catinella,
  %``The velocity function in the local environment from LCDM and LWDM constrained simulations,''
  Astrophys.\ J.\  {\bf 700}, 1779 (2009)
  [arXiv:0906.0585 [astro-ph.CO]].

%\cite{Tikhonov:2008ss}
\bibitem{Tikhonov:2008ss}
  A.~Tikhonov and A.~Klypin,
  %``The emptiness of voids: yet another over-abundance problem for the LCDM model,''
  arXiv:0807.0924 [astro-ph].

%%%%%%%%%%%%%%%%%%%%%%%%%%%%%%%%%%%%%%%%%%%%%%%%%%%%%%%%%%%%%%%%%%%


%\cite{Viel:2013fqw}
\bibitem{Viel:2013fqw}
  M.~Viel, G.~D.~Becker, J.~S.~Bolton and M.~G.~Haehnelt,
  %``Warm Dark Matter as a solution to the small scale crisis: new constraints from high redshift Lyman-alpha forest data,''
  Phys. Rev. D {\bf 88} (4), 043502 (2013)
  [arXiv:1306.2314 [astro-ph.CO]]

%\cite{Maccio':2012uh}
\bibitem{Maccio':2012uh}
  A.~V.~Maccio, O.~Ruchayskiy, A.~Boyarsky and J.~C.~Munoz-Cuartas,
  %``The inner structure of haloes in Cold+Warm dark matter models,''
  Mon.\ Not.\ Roy.\ Astron.\ Soc.\  {\bf 428}, 882 (2013)
  [arXiv:1202.2858 [astro-ph.CO]].

%\cite{Schneider:2013wwa}
\bibitem{Schneider:2013wwa}
  A.~Schneider, D.~Anderhalden, A.~Maccio and J.~Diemand,
  %``Warm dark matter does not do better than cold dark matter in solving small-scale inconsistencies,''
  arXiv:1309.5960 [astro-ph.CO].



%\cite{Hall:1983id}
\bibitem{Hall:1983id}
  L.~J.~Hall and M.~Suzuki,
  %``Explicit R-Parity Breaking in Supersymmetric Models,''
  Nucl.\ Phys.\ B {\bf 231}, 419 (1984).

%\cite{Endo:2013si}
\bibitem{Endo:2013si}
  M.~Endo, K.~Hamaguchi, S.~P.~Liew, K.~Mukaida and K.~Nakayama,
  %``Axino dark matter with R-parity violation and 130 GeV gamma-ray line,''
  Phys.\ Lett.\ B {\bf 721}, 111 (2013)
  [arXiv:1301.7536 [hep-ph]].


%\cite{Planck}
\bibitem{Planck}
  P.~A.~R.~Ade {\it et al.} [Planck Collaboration],
  %``Planck 2013 results. XVI. Cosmological parameters,''
  [arXiv:1303.5076 [astro-ph.CO]].




%\cite{Chun:2004mu}
\bibitem{Chun:2004mu}
  E.~J.~Chun and S.~C.~Park,
  %``Neutrino mass from R-parity violation in split supersymmetry,''
  JHEP {\bf 0501}, 009 (2005)
  [hep-ph/0410242].

%\cite{Choi:1999tq}
\bibitem{Choi:1999tq}
  S.~Y.~Choi, E.~J.~Chun, S.~K.~Kang and J.~S.~Lee,
  %``Neutrino oscillations and R-parity violating collider signals,''
  Phys.\ Rev.\ D {\bf 60}, 075002 (1999)
  [hep-ph/9903465].
  %%CITATION = HEP-PH/9903465;%%
  %80 citations counted in INSPIRE as of 05 Mar 2014




%\cite{Anderhalden:2012qt}
\bibitem{Anderhalden:2012qt}
  D.~Anderhalden, J.~Diemand, G.~Bertone, A.~V.~Maccio and A.~Schneider,
  %``The Galactic Halo in Mixed Dark Matter Cosmologies,''
  JCAP {\bf 1210}, 047 (2012)
  [arXiv:1206.3788 [astro-ph.CO]].


%\cite{Jaeckel:2010ni}
\bibitem{Jaeckel:2010ni}
  J.~Jaeckel and A.~Ringwald,
  %``The Low-Energy Frontier of Particle Physics,''
  Ann.\ Rev.\ Nucl.\ Part.\ Sci.\  {\bf 60}, 405 (2010)
  [arXiv:1002.0329 [hep-ph]].

%\cite{Irastorza:2011gs}
\bibitem{Irastorza:2011gs}
  I.~G.~Irastorza, F.~T.~Avignone, S.~Caspi, J.~M.~Carmona, T.~Dafni, M.~Davenport, A.~Dudarev and G.~Fanourakis {\it et al.},
  %``Towards a new generation axion helioscope,''
  JCAP {\bf 1106}, 013 (2011)
  [arXiv:1103.5334 [hep-ex]].



\bibitem{Raffelt:2006cw}
  G.~G.~Raffelt,
  %``Astrophysical axion bounds,''
  Lect.\ Notes Phys.\  {\bf 741}, 51 (2008)
  [hep-ph/0611350].






%\cite{Visinelli:2014twa}
\bibitem{Visinelli:2014twa}
  L.~Visinelli and P.~Gondolo,
  %``Axion cold dark matter in view of BICEP2 results,''
  arXiv:1403.4594 [hep-ph].

%\cite{BICEP2}
\bibitem{BICEP2}
  P.~A.~R.~Ade {\it et al.}  [BICEP2 Collaboration],
  %``BICEP2 I: Detection Of B-mode Polarization at Degree Angular Scales,''
  arXiv:1403.3985 [astro-ph.CO].



%\cite{Haxton:2000xb}
\bibitem{Haxton:2000xb}
  W.~C.~Haxton and W.~Lin,
  %``The Very low-energy solar flux of electron and heavy flavor neutrinos and anti-neutrinos,''
  Phys.\ Lett.\ B {\bf 486}, 263 (2000)
  [nucl-th/0006055].

\end{thebibliography}
\end{document}